# The Polar Stellar Ring and Dark Halo of NGC 5907

## V. P. Reshetnikov* and N. Ya. Sotnikova

*Astronomical Institute, St. Petersburg State University, Bibliotechnaya pl. 2, Petrodvorets, 198904 Russia*

**Abstract**—Numerical simulations of the disruption of a dwarf companion moving in the polar plane of a massive galaxy are presented. The constructed model is compared with observational data on the recently discovered low-surface-brightness stellar ring around the galaxy NGC 5907. Constraints on the ring lifetime (≤1.5 Gyr after the first approach of the galaxies), on the structure of the companion—the ring precursor, and on the mass of the dark halo of the main galaxy in whose gravitational field the companion moves are provided. The dark-halo mass within 50 kpc of the NGC 5907 center cannot exceed 3 or 4 "visible" masses. © *2000 MAIK "Nauka/Interperiodica"*.

Key words: *galaxies, groups and clusters of galaxies, intergalactic gas*

## 1. INTRODUCTION

In recent years, many new approaches to studying the distribution of mass, in particular, dark mass, in galaxies of various types have appeared because of the progress in observing techniques and computer simulation methods. These include methods based on gravitational lensing [1], on the search for patterns in the variations of the H I layer thickness along the galactic radius [2], on the kinematic study of systems of planetary nebulae [3] and dwarf galaxies [4], as well as on the investigation of the dynamics of precessing dust disks in early-type galaxies [5] and circumgalactic polar structures [6]. One of the most interesting and promising ways of inferring the characteristic parameters of dark halos is to analyze the morphology and kinematics of transient, relatively short-lived structures produced by gravitational interaction between galaxies—tidal tails, bars, and the remnants of companions disrupted by interaction [7–11].

Here, we report the results of our numerical simulations of the large-scale, low-surface-brightness optical feature discovered around the galaxy NGC 5907. By all appearances, this structure results from the disruption of a low-mass companion near a massive spiral galaxy. We discuss the results of our calculations in the context of dark-halo effects on this process.

## 2. THE GALAXY NGC 5907

NGC 5907 is a late-type spiral (Sc) galaxy seen at a large angle to the line of sight ('edge-on') (see Fig. 1a). The galaxy lies relatively nearby (its distance is 11 Mpc for $H_0 = 75$ km s$^{-1}$ kpc$^{-1}$), its optical radius ($R_{opt}$) is 19.3 kpc, and the exponential disk scale is 5.7 kpc [12]. Its absolute magnitude is $M_B = -19.10$ ($L_B = 6.8 \times 10^9$ $L_\odot^B$). The planes of the gaseous and stellar disks of the galaxy exhibit a large-scale warping [13, 14]. Since no close companions of comparative (or even considerably lower) luminosity have been found near NGC 5907, it has long been considered to be the prototype of noninteracting galaxies with warped equatorial planes. Such objects are very rare and are dynamical puzzles, because interaction with the ambient environment is believed to be mainly responsible for the formation of large-scale warping of galactic planes [15].

In recent years, the view of NGC 5907 as an isolated object has been called into question. Sackett *et al.* [16] discovered that the galaxy is surrounded by an extremely faint ($\mu_R \geq 25^m$ arcsec$^{-2}$) optical shell, a halo. The existence of this shell was subsequently confirmed by *B, V,* and *I* observations [17, 18] and *J, H,* and *K* data [19, 20]. Lequeux *et al.* [18] found the color indices of the halo to be typical of the old stellar population of elliptical galaxies. They assumed that the stellar halo around NGC 5907 was produced by the disruption of a relatively low-mass (~$3 \times 10^{10}$ $M_\odot$) elliptical galaxy near it. This assumption was illustrated by numerical calculations.

Shang *et al.* [21] reported the discovery of an even fainter (than the halo) ring-shaped feature around NGC 5907 ($\mu_R \sim 27–28^m$ arcsec$^{-2}$). This feature is similar in shape to an ellipse elongated along the galaxy's minor axis, so that the ring turns out to be in the circumpolar galactic plane. The galaxy nucleus lies approximately at the focus of this ellipse (see Fig. 1b). The major axis of the feature itself (~43 kpc) is comparable to the optical diameter of NGC 5907. The total optical luminosity of the ring does not exceed 1% of the galaxy luminosity, and its color index $R - I \sim 0.9$ corresponds to the old stellar population. No H I line emission was detected from the ring ($M$(HI) $\leq 10^8 \, M_\odot$), nor are there any traces of H$\alpha$ and O[III] emission. The authors of the discovery suggested that the ring-shaped feature was a tidal remnant of the dwarf spheroidal galaxy captured by NGC 5907. Shang *et al.* [21] also discovered a dwarf companion with a radial velocity close to that of the central galaxy within 37 kpc of the NGC 5907 center.

---

* E-mail address for contacts: resh@astro.spbu.ru

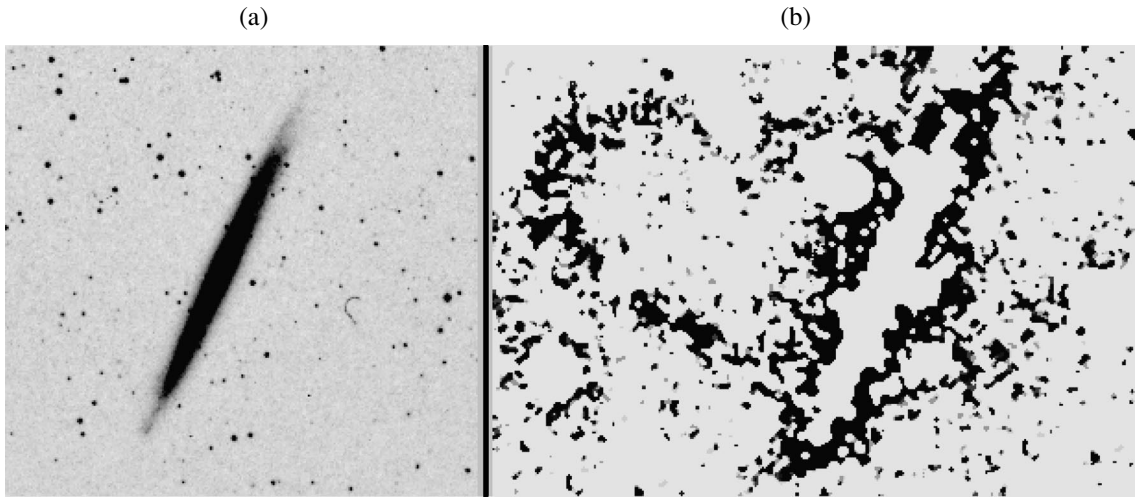

**Fig. 1.** (a) Reproduction of the NGC 5907 image based on the Digital Sky Survey (DSS)[1]; (b) image of the galaxy's outer regions [21]. To enhance the contrast, the bright central region of NGC 5907 and all foreground stars were removed from the image by masking. The image height is $14'.5$; north is at the top, and east is to the left.

It follows from the above data that NGC 5907 can no longer be considered as an isolated galaxy. By all appearances, the peculiarities of its optical structure suggests an ongoing interaction with its low-mass companions.

To test the assumption that the large-scale circumpolar stellar ring is produced by the accretion of a dwarf galaxy onto NGC 5907, we performed a series of numerical simulations, which are describe below.

## 3. MODEL AND COMPUTATIONAL TECHNIQUE

In the computations, the potential of the main galaxy was assumed to be known. As for the companion, we considered two alternative cases: a model of test particles in the external potential of the companion and a self-gravitating companion.

In the first case, we assumed the companion to be disrupted completely at the time of the closest approach of the galaxies, following which the particles continue to move in the field of the massive galaxy alone.

We simulated the evolution of the self-gravitating companion by using the NEMO package [22]. This is a freeware package designed to numerically solve gravitational *N*-body problems. It consists of subroutines for specifying initial configurations of stellar–dynamical systems (including many standard models) and subroutines for simulating the evolution of these systems based on various numerical schemes. In our computations, we used the scheme for constructing an "hierarchical tree" [23]—a special data structure in which all gravitating particles are separated into different groups, depending on the degree of closeness to the point at which the gravitational potential is determined. The contribution from close particles to the potential is calculated by a simple summation, while the groups of distant particles are treated as a single object. The potential produced at a given point by the distant groups of particles is expanded into a series in which the monopole and, occasionally, quadrupole terms are retained. The maximum number of particles used in our numerical simulations is $N = 100\,000$. In this case, we managed to suppress substantially the effects of pair relaxation and to trace the evolution of the disrupted companion on time scales of ~4–5 Gyr.

According to observations (see Sect. 2), the luminosity of the ring-shaped structure formed from the disrupted companion is much lower than that of the main galaxy. We therefore assumed the mass of the companion itself to be also sufficiently small (~$10^{-3}$ of the mass of the main galaxy). This allowed us to ignore dynamic friction and disregard possible energy exchange between the main galaxy and the companion (see, e.g., [24])

### 3.1. The Central Galaxy

For the central galaxy, we considered a model consisting of three components: a disk, a bulge, and a dark halo. The disk potential was represented by Miyamoto–Nagai's potential [25]

---

[1] The Digitized Sky Surveys were produced at the Space Telescope Science Institute under US Government grant NAG W-2166.

**Table 1.** Models of the mass distribution in NGC 5907

| Model | Component | Scale, kpc | Mass within $R_{opt} = 19.3$ kpc, $10^{10} M_\odot$ | $r_c$, kpc |
|---|---|---|---|---|
| 1 | Bulge | 1 | 1.3 | – |
|  | Disk | 5<br>1 | 9.0 | – |
|  | Halo | 20 | 13.5 | 18 |
| 2 | Bulge | 1 | 1.3 | – |
|  | Disk | 5<br>1 | 9.0 | – |
|  | Halo | 20 | 5.0 | ∞ |
| 3 | Bulge | 1 | 1.3 | – |
|  | Disk | 5<br>1 | 9.0 | – |
|  | Halo | 20 | 13.5 | ∞ |

$$\Phi_d(x, y, z) = -\frac{GM_d}{[x^2 + y^2 + (b_d + \sqrt{z^2 + a_d^2})^2]^{1/2}}, \quad (1)$$

where $M_d$ is the disk mass, and $a_d$ and $b_d$ are the parameters characterizing the scale of the mass distribution in the disk.

The galactic bulge was described by Plummer's sphere

$$\Phi_b(r) = -\frac{GM_b}{(r^2 + a_b^2)^{1/2}}, \quad (2)$$

where $M_b$ is the bulge mass, and $a_b$ is the potential-smoothing scale.

The mass distribution in the dark halo of the main galaxy was specified by using the model of a mass-limited isothermal sphere, starting from some distance from the nucleus:

$$M_h(r) = \frac{M_{h0}}{r_c}\left(r - a_h \arctan\left(\frac{r}{a_h}\right)\right)\left(1 - e^{-\frac{r_c}{r}}\right), \quad (3)$$

where $M_{h0}$ is the total mass of the halo, and $a_h$ is the radius of the halo nucleus. The parameter $r_c$ was introduced to avoid the divergence of mass at large distances in the isothermal model. Up to distances of the order of $r_c$, the halo mass density varies as $\propto r^{-2}$, while the mass $M(r)$ increases as $\propto r$. Starting from $r_c$, the density drops abruptly, and the mass tends to a finite value of $M_{h0}$.

In a number of simulations, we did not introduce a restriction by mass. In this case, the density distribution in the isothermal sphere was taken in ordinary form

$$\rho_h(r) = \frac{\rho_{h0}}{1 + (r/a_h)^2}. \quad (4)$$

The model parameters were chosen in such a way that the rotation curve within the optical radius (19.3 kpc) was close to the observed one. The model of mass distribution in NGC 5907 we used is similar to that described in [26] (see also [18]). Table 1 gives parameters of the main galaxy for our three cases. In all cases, the rotation curve has a flat portion in the range 8 to 30 kpc (Fig. 2).

### 3.2. The Companion

We considered two companion models (see Table 2). One of them is Plummer's spherically-symmetric model

$$\Phi_s(r) = -\frac{GM_s}{(r^2 + a_s^2)^{1/2}}, \quad (5)$$

where $M_s$ is the companion mass, and $a_s$ is the scale length of the mass distribution.

As was already noted above, the choice of companion mass [$(2–5) \times 10^8 M_\odot$] was constrained by the observed optical luminosity of the ring, which does not exceed 1% of the NGC 5907 luminosity. For the above companion masses and the total luminosity of the feature under consideration ($\leq 7 \times 10^7 L_\odot^B$), we found the mass-to-light ratio to be $M_s/L_s(B) \geq 3–7$, in agreement with data for actual dwarf galaxies [27].

Table 2 gives line-of-sight velocity dispersions $\sigma_0 = [GM_s/6a_s]^{1/2}$ for the models of dwarf spheroidal companions. These values also agree with the observed $\sigma_0$ for dwarf spheroidal galaxies (see, e.g., the parameters of NGC 147 and NGC 185 in the review [27]).

In the second case, the companion was represented by a disk galaxy with an exponential density distribu-tion and a vertical-to-radial scale ratio of 1 : 5 (Table 2). The Toomre parameter was set equal to $Q_T = 1.5$.

**Table 2.** Models of the companion

| Disk companion | Exponential scale, kpc | Vertical scale, kpc | Mass, $10^8 M_\odot$ |
|---|---|---|---|
|  | 0.4 | 0.08 | 2<br>5 |
| Spheroidal companion | Scale ($a_s$) | $\sigma_0$, km s$^{-1}$ | Mass, $10^8 M_\odot$ |
|  | 0.4 | 19<br>30 | 2<br>5 |

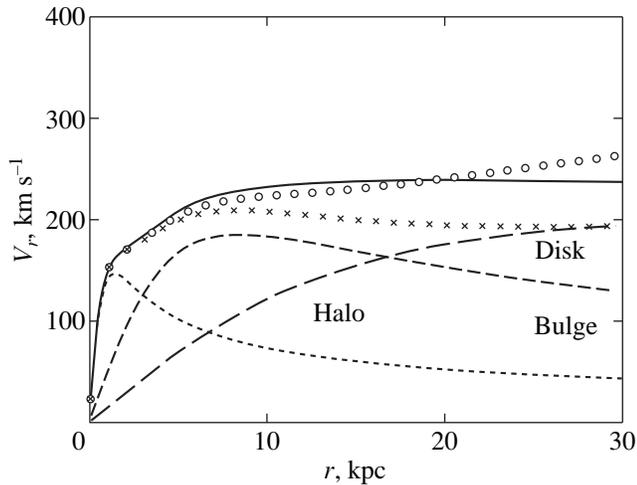

**Fig. 2.** Rotation curve of NGC 5907. The solid line indicates the rotation curve for model 1, and the dashed lines indicate the contributions of different components. The rotation curves for models 2 and 3 are represented by crosses (×) and circles (○), respectively.

### 3.3. The Orbit

The companion was assumed to move in the polar plane of the main galaxy (*xz* plane). The initial coordinates of the companion are $x_0 = 0$, $y_0 = 0$, and $z_0 = 52$ kpc. We chose the initial velocity from the following considerations. We specified the distance of the closest approach, $r_{min} = 13$ kpc, and computed the galaxy mass $M(r_{min})$ within $r_{min}$. If the companion moved in the field of a point mass $M(r_{min})$, then its velocity at point $(x_0, y_0, z_0)$ would correspond to the velocity at the apocenter of an elliptical orbit with eccentricity $e = 0.6$ and semimajor axis $a = (z_0 + r_{min})/2 = 32.5$ kpc. It is this value that we took as the initial velocity. Figure 3 shows the companion orbit for the three models of the main galaxy. The distance of the closest approach proved to be less than the initially specified value (~8–9 kpc) for all three orbits.

For the disk companion, we considered three different initial orientations of the disk plane with respect to the orbital plane and the equatorial plane of the galaxy:

(a) The disk plane is parallel to the galaxy's equatorial plane;

(b) The disk plane is perpendicular both to the galaxy's equatorial plane and to the orbital plane (in the orbital plane, the companion is seen edge-on);

(c) The disk companion moves edge-on with respect to the galaxy plane, with the disk being seen face-on in the orbital plane.

Our computations revealed virtually no differences between the direct and retrograde motions of the disk companion. This is because the stars at the opposite ends of the disk undergo virtually the same perturbation due to the small size of the companion itself. Thus, the difference between the perturbation durations for the direct and retrograde rotation of the companion is of no importance. Below, we report the results of our computations only for direct motions.

## 4. RESULTS OF COMPUTATIONS

### 4.1. Test Particles

We assumed that the companion was completely disrupted when the separation between the galaxies was at a minimum. Subsequently, its remnants continue to move only in the gravitational field of the main galaxy. In the course of time, the companion remnants stretch along the initial orbit following its individual turns. We traced the evolution of the disrupted companion within 1.4 Gyr after the first closest approach of the galaxies. Figures 4a and 4b show the results of our computations for the two (spherical and disk) companion models. The spatial orientation of the disk companion corresponds to case (b) in Subsect. 3.3. The main galaxy is represented by model 1 (Table 1).

We see that a diffuse loop-shaped structure resembling that in Fig. 1b forms around the galaxy as the companion is disrupted (by $t \sim 1.4$ Gyr after the first closest approach). Note also that the numerical model predicts the existence of an extended linear feature going off the simulated loop to the left and downward (Fig. 4). NGC 5907 also shows evidence of such a structure (Fig. 1b). Since the companion remnants move exactly in the polar plane, the loop is not washed out by differential precession. For the disk companion, the loop is thinner because of the smaller stellar velocity dispersion along the *x* axis.

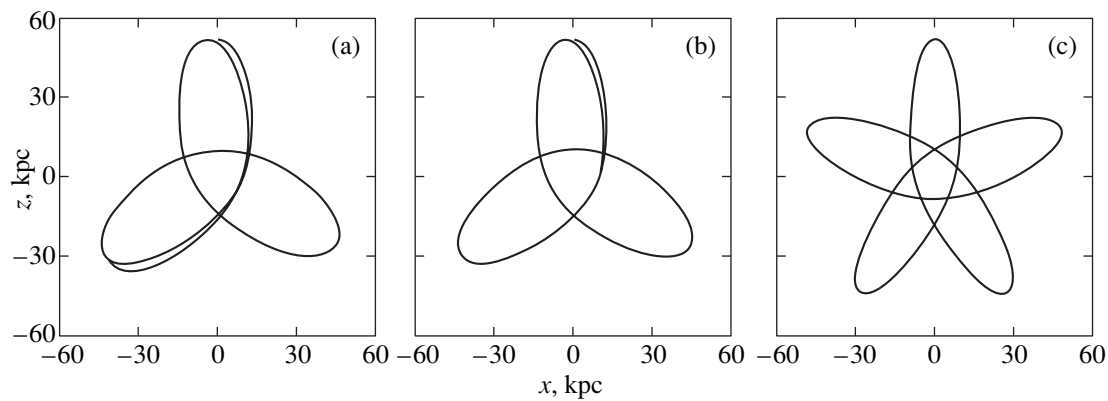

**Fig. 3.** The orbit of the companion for models 1 (a); 2 (b); and 3 (c).

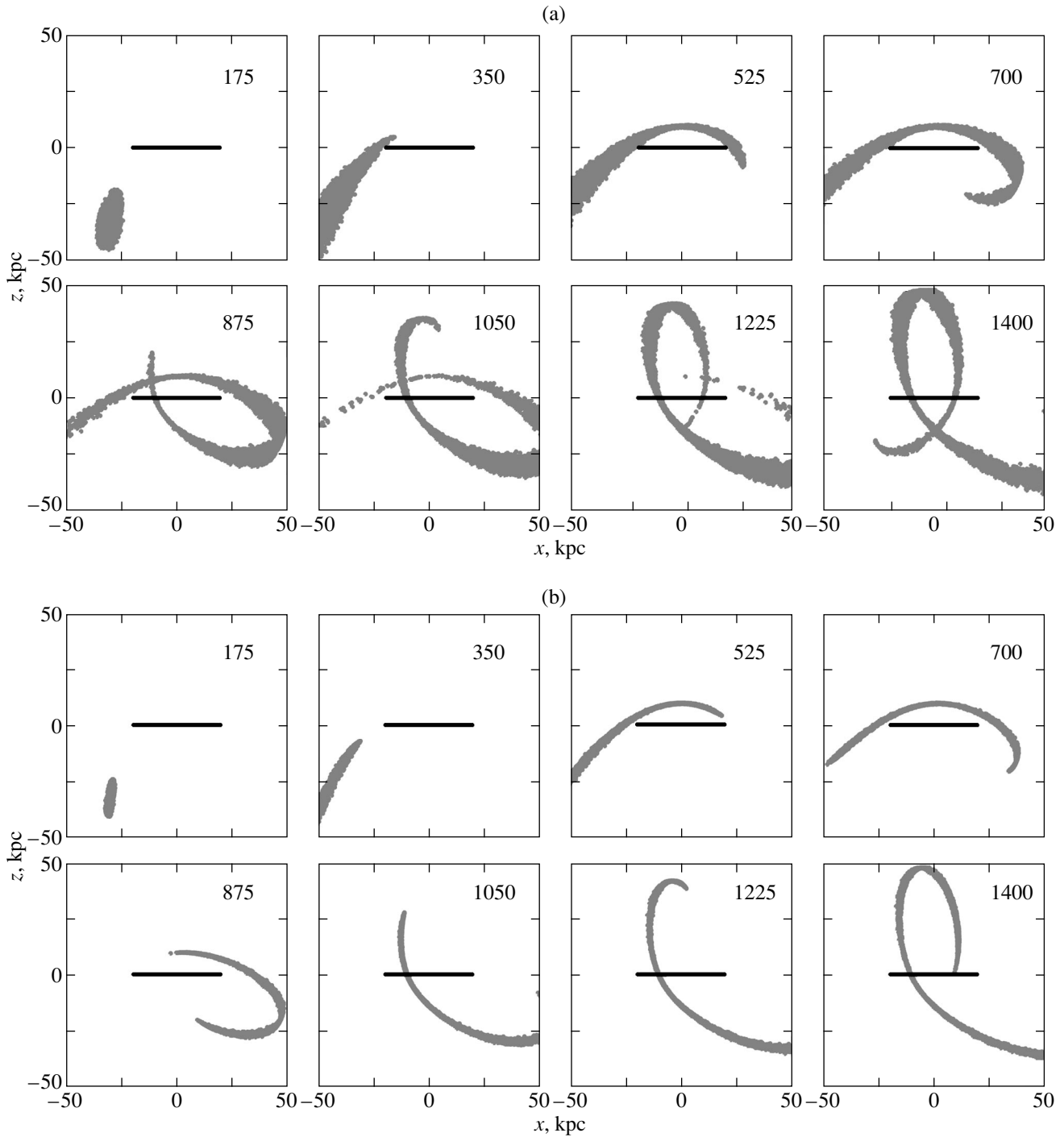

**Fig. 4.** Evolution of the structure of the dwarf companion in the model of test particles in projection onto the *xz* plane: (a) spherical companion and (b) disk companion. The disk companion was initially located in the *yz* plane. The companion mass is $2 \times 10^8 \, M_\odot$ in both cases. A straight-line segment indicates the orientation of the disk of the main galaxy. The segment length corresponds to the optical diameter of NGC 5907. The size of each frame is $100 \times 100$ kpc; the time (in Myr) elapsed since the first passage of the companion through the pericenter of the orbit is shown in the upper right corner.

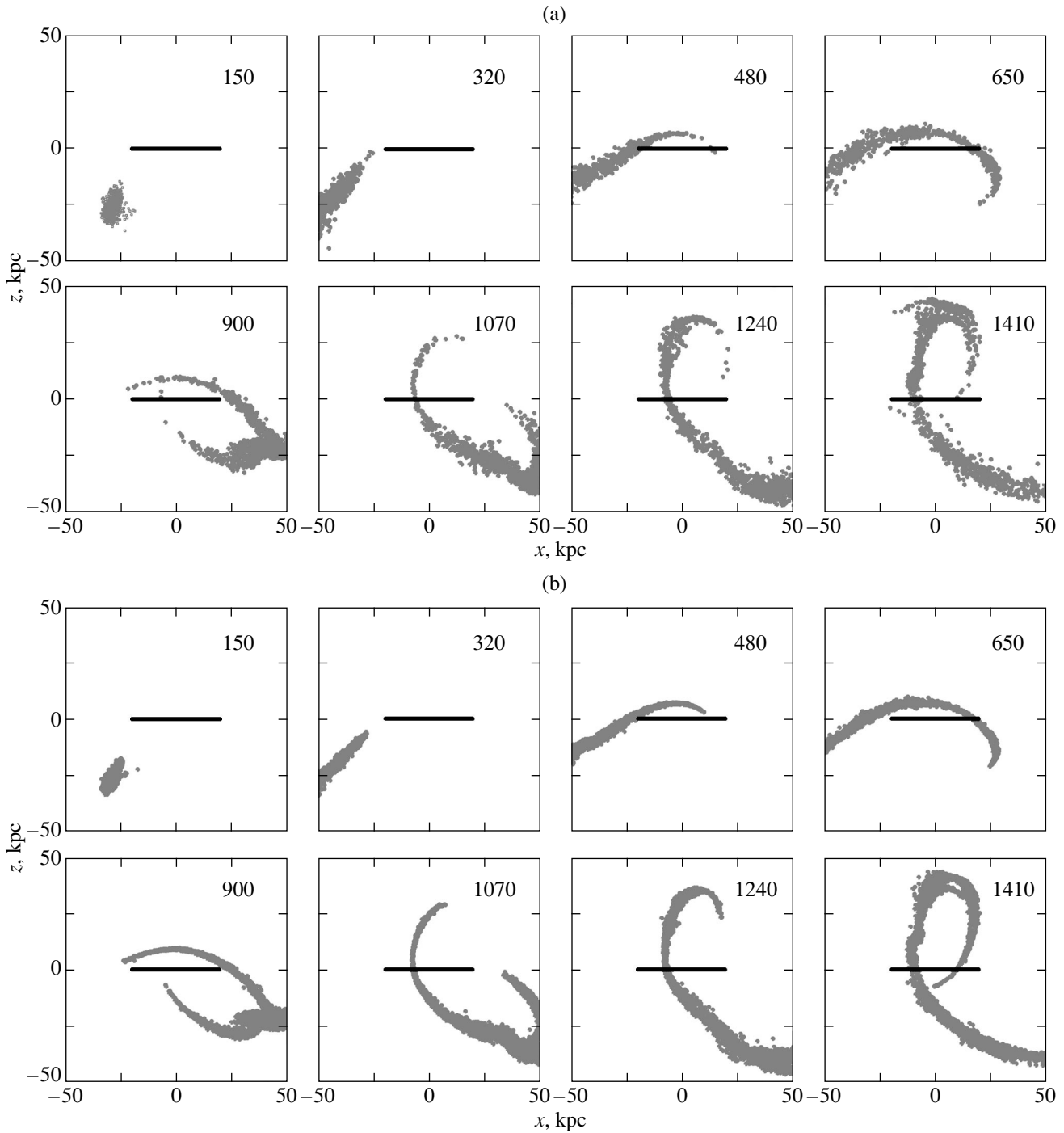

**Fig. 5.** Same as Fig. 4 but for the model of a self-gravitating companion.

In model 2 (Table 1), the potential of the halo corresponds to the potential of a standard isothermal sphere with infinite total mass. In order that the total mass of dark matter within the region of companion motion (~52 kpc) be approximately the same as that in model 1, we reduced the halo mass within $R_{opt}$. Our computations show that a structure similar to that in Figs. 4a and 4b is also formed in this case.

It also follows from our computations that model 3 (Table 1) is unacceptable. In this model, the mass of the halo within the region of companion motion is considerably higher than that in models 1 and 2. We see from Fig. 3 that, in this case, the apparent flattening of the turns of the orbit along which the disrupted companion stretches is too small compared to what is observed in the ring-shaped structure of NGC 5907. Consequently, even the morphology itself of the ring-shape feature constrains the mass of the dark halo of NGC 5907 within the ring, $M_h (r < 52 \text{ kpc})/(M_d + M_b) \approx 3\text{--}4$.

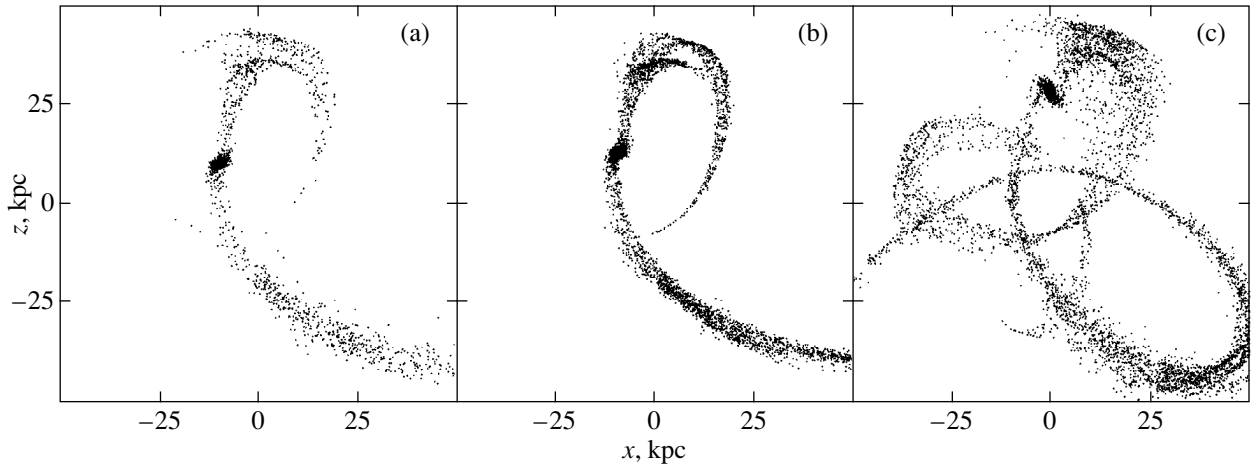

**Fig. 6.** Morphology of the ring produced by the companion disruption: (a) the model of a spherical companion ($t = 1.41$ Gyr); (b) early ($t = 1.41$ Gyr) and (c) late ($t = 3.51$ Gyr) stages of the ring evolution for the model of a disk companion.

### 4.2. The Self-Gravitating Companion

To describe the companion disruption in a self-consistent way, we computed several models in the gravitational $N$-body problem. Figures 5a and 5b show the evolution of the spherical and disk companions. We used $N = 50\,000$ and $N = 100\,000$ in the former and latter cases, respectively. It is not accidental that we took such a large number of particles. We wished to trace the evolution of the companion on time scales of 2–4 Gyr. For a smaller number of particles, this evolution is largely determined by numerical relaxation effects, which cause artifical heating of the system. The relaxation time $t_{rh}$ for a companion with the potential (5) is given by [28]

$$t_{rh} = 0.2 \frac{N}{\ln(0.4N)} \sqrt{\frac{a_s^3}{GM_s}}. \qquad (6)$$

For $N = 50000$, $M_s = 2 \times 10^8 M_\odot$, and $a_s = 0.4$ kpc, this yields about 8 Gyr.

Qualitatively, the formation of the stellar ring is similar in pattern to that described in Subsect. 4.1. The principal difference is that, even 1.4 Gyr after the galaxies approached each other for the first time, the nucleus of the companion proves to be virtually undisrupted (see Figs. 6a and 6b). By that time, the companion had crossed four times the equatorial plane of the main galaxy. Note that the disk companion is disrupted to a greater extent than the spherical one. Accordingly, the surface density of the ring formed from the remnants of the disk companion (0.1–0.5 $M_\odot$ pc$^{-2}$) is considerably (approximately a factor of 3) higher than that in the case shown in Fig. 6a. Such a density is in good agreement with the observed surface brightness of the ring (Sect. 2).

Figure 6c shows the late disruption stage of the disk companion (after eight crossings of the galaxy's equatorial plane). A small nucleus is still bound, but the companion remnants no longer form a coherent ring-shaped feature but are distributed along several turns of the orbit. Similar computations by other authors (see, e.g., [29]) on long time scales also yielded highly irregular structures. This constrains the lifetime of the stellar loop observed in NGC 5907 (< 1.5 Gyr).

Note that the behavior of stars differs from the behavior of gas. The gas captured by a massive galaxy forms a long-lived ring, which can give rise to objects called polar-ring galaxies [29–31]. The fate of the stellar ring in NGC 5907 is different: in the course of time, it must turn into a diffuse, virtually unobservable structure.

## 5. CONCLUSION

Our computations have led us to several conclusions:

(1) The ring-shaped feature discovered around NGC 5907 is satisfactorily described as the remnant of a disrupted low-mass (~$10^{-3}$ of the mass of the main galaxy) companion.

(2) The structure of the stellar loop can be better explained by assuming that the dwarf companion was a disk rather than a spherical one.

(3) In NGC 5907, we are apparently observing a relatively early stage of companion disruption (<1.4 Gyr after the first closest approach of the galaxy and its companion). At late stages, the companion matter is distributed simultaneously along several turns of the orbit.

(4) The ring morphology allows the mass of the dark halo of the main galaxy within the ring radius to be constrained: $M_h(r < 52 \text{ kpc})/(M_d + M_b) \approx 3\text{–}4$.

(5) Deep images of galaxies can apparently reveal faint features in some of them similar to the stellar ring around NGC 5907. Our Galaxy is one of such examples. Its dwarf companion in Sagittarius, which was partly disrupted by tidal forces from the Galaxy, forms an extended structure (with a length of 68° or ~30 kpc [32]) consisting of moving stellar groups (the results of numerical simulations are presented in [33]). However, it would be unreasonable to expect the detection of a large number of such loop-shaped features. First, a special encounter geometry is required for a stellar ring to be formed in the polar plane. Second, in contrast to classical polar-ring galaxies, a stellar ring is observed only when the companion matter is distributed along one of the trajectory turns. The structure becomes blurred at late stages and can hardly be detected, because the surface brightness decreases sharply.

## ACKNOWLEDGMENTS

We are grateful to Dr. Shang (Texas University), who provided a deep image of NGC 5907 (Fig. 1b). This study was supported by the Integration Program (project no. 578), as well as by the Competitive Center for Basic Science of the Ministry of General and Professional Education and the Russian Foundation for Basic Research (project nos. 97-02-18212 and 98-02-18178).

*Translated by A. Dambis*